\documentclass[fleqn,12pt,twoside]{article}

\usepackage[headings]{espcrc1}
\readRCS
$Id: espcrc1.tex,v 1.2 2004/02/24 11:22:11 spepping Exp $
\usepackage{graphicx}

\usepackage[figuresright]{rotating}

\newtheorem{theorem}{Theorem}

\newtheorem{conjecture}{Conjecture}
\newtheorem{corollary}{Corollary}

\newtheorem{definition}{Definition}

\newtheorem{lemma}{Lemma}

\newtheorem{problem}{Problem}

\newtheorem{remark}{Remark}

\newenvironment{dedication}[1][]{\begin{trivlist}
\item[\hskip \labelsep {\bfseries #1}]}{\end{trivlist}}

\newenvironment{proof}[1][Proof.]{\begin{trivlist}
\item[\hskip \labelsep {\bfseries #1}]}{\end{trivlist}}

\newenvironment{acknowledgement}[1][Acknowledgement]{\begin{trivlist}
\item[\hskip \labelsep {\bfseries #1}]}{\end{trivlist}}

\newcommand{\AmS}{{\protect\the\textfont2
  A\kern-.1667em\lower.5ex\hbox{M}\kern-.125emS}}

\usepackage{amsmath}
\usepackage{amsfonts}
\title{The hardness of the independence and matching clutter of a graph}

\author{Sasun Hambartsumyan \address[MCSD]{Department of Informatics and Applied Mathematics,\\
Yerevan State University, Yerevan, 0025, Armenia}\thanks{email:
hsasun@yahoo.com },
Vahan V. Mkrtchyan\addressmark[MCSD]%
\address{Institute for Informatics and Automation Problems,\\
National Academy of Sciences of Republic of Armenia, 0014,
Armenia}\footnote{The authors are supported by a grant of Armenian
National Science and Education Fund \label{ANSEF}}\thanks{email: vahanmkrtchyan2002@\{ysu.am, ipia.sci.am,
yahoo.com\}},
        Vahe L. Musoyan\addressmark[MCSD]\address{Computer Science
        Department,\\
Stanford University, Stanford, CA 94305,
USA}\thanks{email: vmusoyan@cs.stanford.edu,
vahe.musoyan@gmail.com}\ref{ANSEF},
                and
        Hovhannes Sargsyan \addressmark[MCSD]\thanks{email: hsargsian@gmail.com }}

\runtitle{The hardness of the independence and matching clutter of a graph}
\runauthor{S. Hambartsumyan, V. V. Mkrtchyan, V. L. Musoyan, H.
Sargsyan}

\begin{document}

% typeset front matter
\maketitle

\begin{abstract}
A {\it clutter} (or {\it antichain} or {\it Sperner family}) $L$ is a pair $(V,E)$, where $V$ is a finite set and $E$ is a family of subsets of $V$ none of which is a subset of another. Usually, the elements of $V$ are called {\it vertices} of $L$, and the elements of $E$ are called {\it edges} of $L$. A subset $s_e$ of an edge $e$ of a clutter is called {\it recognizing} for $e$, if $s_e$ is not a subset of another edge. The {\it hardness} of an edge $e$ of a clutter is the ratio of the size of $e\textrm{'s}$ smallest recognizing subset to the size of $e$. The hardness of a clutter is the maximum hardness of its edges. We study the hardness of clutters arising from independent sets and matchings of graphs.
\end{abstract}\\ 



\begin{dedication}
$$\textit{Dedicated to Professor Stepan E. Markosyan}$$
\end{dedication}

\section{Introduction}

A {\it clutter} (or {\it antichain} or {\it Sperner family}) $L$ is a pair $(V,E)$, where
$V$ is a finite set and $E$ is a family of subsets of $V$ none of
which is a subset of another. Following \cite{Cornuejols}, the
elements of $V$ will be called {\it vertices} of $L$, and the elements
of $E$ are called {\it edges} of $L$.

Given a clutter $L=(V,E)$, a subset $e_0\subseteq e$ of an edge $e$ is a
{\it recognizing} subset for $e$, if $e_0\subseteq e'$ for some $e'\in E$, then $e'=e$. Let $S_e$ be a smallest recognizing subset of $e\in E$, $c(e)=|S_{e}|/|e|$, and
\begin{equation*}
c(L)=\max_{e \in E}\textrm{ }c(e).
\end{equation*}

$c(L)$ is called the {\it hardness} of $L$. Note that $0\leq c(L)\leq 1$ for any clutter $L=(V,E)$. Moreover, if $|E|\leq 1$, then clearly $c(L)=0$. Thus, it is natural to consider clutters $L$ with at least two edges. In this case any edge contains no more than $|V|-1$ vertices and any recognizing subset of an edge $e\in E$ must contain at least one vertex. Thus,
\begin{equation*}
\frac{1}{|V|-1}\leq c(L)\leq 1,
\end{equation*}and the lower bound is tight, since if for any positive integer $n$ ($n\geq 2$) we take $V_n=\{1,...,n\}$, $E_n=\{\{1,3,...,n\},\{2,3,...,n\}\}$, then clearly $L_n=(V_n,E_n)$ is a clutter, the sets $\{1\}$ and $\{2\}$ are recognizing subsets for the edges $\{1,3,...,n\}$ and $\{2,3,...,n\}$, respectively, and
\begin{equation*}
c(L_n)=\frac{1}{n-1}=\frac{1}{|V_n|-1}.
\end{equation*} Note that the main reason why the clutter $L_n$ has such a low hardness, is that the elements $3,...,n$ are present in every edge. This means that they cannot be present in any smallest recognizing subset of an edge, and therefore they do not contribute to the numerator of the hardness of an edge, however, they do contribute to its denominator. This situation prompts the following
\begin{problem}\label{MainProblem} Find a best-possible function $f$ such that any clutter $L=(V,E)$ satisfying the condition
\begin{enumerate}
	\item [(C1)] no vertex of $L$ is present in all edges of $L$,
\end{enumerate}has hardness $c(L)\geq f(|V|)$.
\end{problem}Our considerations above imply that $f(|V|)\geq \frac{1}{|V|-1}$. Let us also note that any clutter satisfying (C1) has at least two edges, moreover, without loss of generality, we can assume that the clutters in the formulation of Problem \ref{MainProblem} satisfy
\begin{enumerate}
	\item [(C2)] each vertex of $L$ is present in at least one edge of $L$,
\end{enumerate}since isolated vertices (vertices, that do not belong to an edge) can be removed from the clutter without affecting its hardness.

In this paper, we address the Problem \ref{MainProblem} for two classes of clutters that arise from graphs. Let us note that the graphs considered in this paper are finite, undirected and do
not contain multiple edges or loops. Formally, such a graph $G$ can be considered as a clutter $(V,E)$, in which $E$ is any subset of the set of pairs of elements from $V$.

For a graph $G$, let $V(G)$ and
$E(G)$ be the sets of vertices and edges of $G$, respectively. There is an important comment that should be made here concerning the terminology. An edge of a clutter is a subset of the set of vertices, and therefore it can contain more than two vertices, however, an edge of a graph contains exactly two vertices. 

For a vertex $v\in V(G)$ let $d(v)$ be the {\it degree} of $v$, and
let $\Delta(G)$ be the {\it maximum degree} of a vertex of $G$. If
$E\subseteq E(G)$, then let $V(E)$ be the set of vertices of $G$,
which are incident to an edge from $E$. For $S\subseteq V(G)$ let
$G[S]$ denote the subgraph of $G$ {\it induced} by the set $S$.

If $u,v$ are vertices of a graph $G$, then let $\rho(u,v)$ denote
the {\it distance} between the two vertices, that is, the length of the shortest path connecting vertices $u$ and $v$. Moreover, let $diam(G)$ denote the
{\it diameter} of $G$, that is, the maximum distance among any two vertices of $G$.

For a positive integer $n$ let $K_{n}$ denote the complete graph on
$n$ vertices. If $m$ and $n$ are positive integers, then assume
$K_{m,n}$ to be the complete bipartite graph one side of which has
$m$ vertices and the other side has $n$ vertices.

A set $V'\subseteq V(G)$ is said to be {\it independent}, if $V'$ contains no adjacent vertices. Similarly, $E'\subseteq E(G)$ is independent, if $E'$ contains no adjacent edges. An independent set of vertices (edges) is called {\it maximal}, if it does not lie in a larger independent set. An independent set of edges is also called {\it matching}.

Independent sets give rise to clutters. If for a graph $G=(V,E)$ we denote the set of all maximal independent sets of vertices of $G$ by $U_{G}$, then $(V,U_{G})$ is a
clutter. In the paper we use $\mathcal{U}_G$ to denote the clutter $(V,U_G)$. We will need the following properties:
\begin{enumerate}
	\item [(P1)] Any independent set of vertices (particularly, a vertex) of a graph $G$ can be extended to a member of $U_{G}$.
\end{enumerate} Note that if a vertex $v$ belongs to a member $U_v$ of $U_{G}$, then the neighbours of $v$ are not in $U_v$. This implies:
\begin{enumerate}
	\item [(P2)] In any graph $G$, $\min_{U \in U_{G}}|U|+ \Delta(G) \leq|V(G)|$.
\end{enumerate} If a graph $G$ contains at least one edge, then $U_{G}$ contains at least two maximal independent sets. Hence, the empty set is not a recognizing subset. Consider a smallest maximal independent set $U_0$ from $U_{G}$. (P2) implies that $|U_0|\leq |V(G)|-\Delta(G)$. Since the empty set is not a recognizing subset for $U_0$, we have that $c(U_0)\geq \frac{1}{|V(G)|-\Delta(G)}$. Thus,
\begin{enumerate}
	\item [(P3)] If $|E(G)|\geq 1$, then $c(\mathcal{U}_G)\geq \frac{1}{|V(G)|-\Delta(G)}$.
\end{enumerate}

Another clutter that a graph $G=(V,E)$ gives rise is $(E,M_G)$, where $M_G$ denotes the set of all maximal matchings of $G$. This clutter will be denoted by $\mathcal{M}_G$.

The aim of this paper is the investigation of $c(\mathcal{U}_G)$ and $c(\mathcal{M}_G)$. In Theorem \ref{MainBound} in Section 2, we show that 
\begin{equation*}
c(\mathcal{U}_G)\geq \frac{1}{1+|V(G)|-2\sqrt{|V(G)|-1}}
\end{equation*}
provided that $G$ is a connected graph different from $K_1,K_{2,2},K_{3,3},K_{4,4}$. Moreover, we show that this bound is attained by infinitely many graphs. Note that this implies that in the search of the function $f$ for Problem \ref{MainProblem}, one should restrict herself/himself exclusively to those functions that satisfy the following inequality:
\begin{equation*}
\frac{1}{|V|-1}\leq f(|V|)\leq \frac{1}{1+|V|-2\sqrt{|V|-1}}.
\end{equation*} The following example shows that the last inequality is not sharp. Let $k$ be any positive integer with $k\geq 2$. Take $n=k^2$, and let $U_0$ be a set with $n-k=k(k-1)$ elements. Consider an $n$-vertex graph $G$ obtained from a $k$-clique $Q$, by joining every vertex of $Q$ to $k-1$ elements of $U_0$ (each element of $U_0$ is joined to exactly one vertex of $Q$). Note that $U_0\in U_G$. Let $L$ be the clutter  that is obtained from $\mathcal{U}_G$ by removing the edge $U_0$. Observe that all edges of $L$ contain exactly $1+(k-1)^2$ vertices, moreover, a set comprised of a vertex of $Q$ is a smallest recognizing subsets for an edge of $L$. Thus
\begin{equation*}
c(L)=\frac{1}{1+(k-1)^2}.
\end{equation*}It is routine to verify that since $k\geq 2$, we have:
\begin{equation*}
c(L)<\frac{1}{1+|V|-2\sqrt{|V|-1}}.
\end{equation*}
In Theorem \ref{AllRationals} we show that any rational number between $0$ and $1$ is a hardness of a clutter $\mathcal{U}_G$ for some graph $G$. In the end of the Section 2, we make an attempt to characterize the class of trees $T$, for which $c(\mathcal{U}_T)=1$. Though we fail to do this, we present some necessary and some sufficient conditions. We close the section by giving some examples of trees, which show that our conditions are merely necessary or sufficient.

In Section 3, we investigate the hardness of clutters $\mathcal{M}_G$. In a direct analogy with Theorem \ref{AllRationals}, we show that any rational number between $0$ and $1$ is a hardness of a clutter $\mathcal{M}_G$ for some graph $G$. Theorem \ref{ComplexityBoundMG} offers a tight and a better bound for $c(\mathcal{M}_G)$, than one can derive from Theorem \ref{MainBound}. And finally, in the end of the section we investigate the hardness of clutters $\mathcal{M}_G$ arising from regular graphs.

The final Section 4 is devoted to the investigation of some computational problems that are intimately related to the algorithmic computation of $c(\mathcal{U}_G)$. Our investigations show that these problems are $NP$-hard. Let us note that we failed to achieve similar results for the clutters $\mathcal{M}_G$.

Data compression provides a suitable language for the explanation of the essence of our hardness. Suppose that we want to save a maximal matching $H$ of a graph $G$. Of course, it does not make sense for us to save the whole matching $H$. We can keep only its smallest recognizing subset $H_S$, as the set $H\backslash H_S$ is
unique and it can be easily reconstructed from $H_S$. Clearly, $c(\mathcal{M}_G)$ shows the relative hardness of the "worst" maximal matching of $G$.

The hardness of a clutter, that we introduce in the paper, is new (see \cite{Claus,Jukna} where the authors introduce two different types of hardness for graphs). Terms and concepts that we do not define can be found in
\cite{Cornuejols,Lov,West}.

\section{The hardness of $\mathcal{U}_G$}

We start with a lemma, which for a fixed $U\in U_{G}$ gives a necessary and sufficient condition for a set to be recognizing for $U$.

\begin{lemma}\label{recognizingCharacter} Let $U\in U_{G}$. A set $U_{0}\subseteq U$ is recognizing for $U$, if and only if each vertex $v\in V(G)\backslash U$
has a neighbour in $U_{0}$.
\end{lemma}
\begin{proof}Necessity. Assume that $U_{0}$ is recognizing for $U$. Let us show that each vertex lying outside $U$ has a neighbour in $U_0$. Suppose that there is a vertex $v\in V(G)\backslash
U$ that has no neighbour in $U_{0}$. Then, due to (P1), there is $U'\in U_{G}$ such that $U_{0}\cup \{v\}\subseteq U'$. Note that $U'\neq U$ since $v \notin U$. Taking
into account that $U_{0}\subseteq U$ and $U_{0}\subseteq U'$, we
deduce that the set $U_{0}$ is not recognizing for $U$, which contradicts our assumption.

Sufficiency. Now assume that each vertex lying outside $U$ has a neighbour in $U_0$. Let us show that $U_{0}$ is recognizing for $U$. Suppose that the set $U_{0}$ is not recognizing for
$U$. Then there is $U'\in U_{G}$, $U'\neq U$ such that
$U_{0}\subseteq U'$. Since $U'\neq U$, there is $v\in U'\backslash
U$. Note that the vertex $v$ has no neighbour in the set $U_{0}$.
Contradiction. $\square$
\end{proof}

\begin{corollary}\label{specificVertex}If $U\in U_{G}$ and there is
a vertex $v\in V(G)\backslash U$ that has only one neighbour $u$ in
the set $U$, then all recognizing sets of $U$ contain the vertex
$u$.
\end{corollary}

Our next result gives some structural properties of connected graphs $G$, for which any smallest recognizing set of $U\in U_{G}$ has exactly one vertex.

\begin{lemma}\label{SrtuctProperties} Let $G=(V,E)$ be a connected
graph such that all smallest recognizing sets of members of $U_{G}$ contain one vertex. Then:
\begin{enumerate}
\item[(a)] for each $U\in U_{G}$ and its smallest recognizing set $S_U$, the vertex from $S_{U}$ is adjacent to
all vertices outside $U$;

\item[(b)] $\min_{U \in U_{G}}|U|+ \Delta(G) =|V(G)|;$

\item[(c)] Suppose that $U_{G}=\{U_{1},...,U_{l}\}$. Define $S_{G}=\{v\in V(G): v$ lies in exactly one $U \in U_{G}\}$, and for $i=1,...,l$ let $S_{G}(U_{i})=\{x\in S_{G}:x \in
U_{i}\}$. Then any $l$ vertices $u_{1},...,u_{l}$ with $u_{i}\in
S_{G}(U_{i})$ induce a maximum clique of $G$. Moreover, every
maximum clique of $G$ can be obtained in this way;

\item[(d)] $diam(G) \leq3$.
\end{enumerate}
\end{lemma}
\begin{proof}(a) directly follows from Lemma
\ref{recognizingCharacter}.

(b) Choose $U_{0}\in U_{G}$ with $|U_{0}|= \min_{U \in U_{G}}|U|$.
According to (a), there is an $x\in U_{0}$ that is adjacent to all
vertices from $V(G)\backslash U$. Note that
\begin{equation*}
\Delta(G)\geq d(x)=|V(G)|-|U_{0}|=|V(G)|-\min_{U \in U_{G}}|U|,
\end{equation*}
thus
\begin{equation*}
\Delta(G)\geq |V(G)|-\min_{U \in U_{G}}|U|.
\end{equation*}
(P2) implies that
\begin{equation*}
\Delta(G)+\min_{U \in U_{G}}|U|=|V(G)|.
\end{equation*}

(c) Let $U_{i}\in U_{G}$ and $U_{j}\in U_{G}$ ($i\neq j$), and
consider vertices $v_{i}\in S_{G}(U_{i})$ and $v_{j}\in
S_{G}(U_{j})$. Clearly, $v_{i}\notin U_{j}$ and $v_{j}\notin U_{i}$,
hence due to (a) $(v_{i},v_{j})\in E(G)$. This implies that any vertices
$u_{1},...,u_{l}$ with $u_{i}\in S_{G}(U_{i}),i=1,...,l$ induce a
clique of $G$, and particularly, the size of the maximum clique of
$G$ is at least $l$.

Thus to complete the proof of (c), we only need to show
that for any maximum clique $Q$ of the graph $G$ there are
$u_{1},...,u_{l}$ with $u_{i}\in S_{G}(U_{i}),i=1,...,l$, such that
$V(Q)=\{u_{1},...,u_{l}\}$.

Let $Q$ be a maximum clique of the graph $G$, and let $U\in U_{G}$.
Clearly, $|V(Q)\cap U|\leq 1$. Let us show that $|V(Q)\cap U|=1$. If
$V(Q)\cap U=\emptyset$, then due to (a), there is $x \in U$ such that
$x$ is adjacent to all vertices of $Q$. This implies that the set $V(Q)\cup
\{x\}$ forms a larger clique of $G$ contradicting the choice of $Q$.

Thus $|V(Q)\cap U|=1$. Suppose that $V(Q)\cap U=\{x\}$. Let us show
that $x\in S_{G}(U)$. Suppose not. Then there is $U'\in U_{G},U'\neq U$ such that 
$x\in U'$. Clearly, $V(Q)\cap U'=\{x\}$. Let $u\in U$ and $u'\in U'$ be vertices such that $\{u\}$ and $\{u'\}$ are recognizing subsets for $U$ and $U'$, respectively. (a) implies
that the vertices $u$ and $u'$ are adjacent to
all vertices lying outside $U$ and $U'$, respectively. Since $x\in
U,U'$, we imply that $u$ and $u'$ do not belong to the clique $Q$.
 Now, it is not
hard to see that the set $(V(Q)\backslash\{x\})\cup \{u,u'\}$
induces a clique that is larger than $Q$ contradicting the choice of
$Q$. Thus $x\in S_{G}(U)$ and the proof of (c) is completed.

(d) Suppose that $diam(G) \geq4$, and consider the vertices $u,v\in
V(G)$ with $\rho(u,v)=diam(G) \geq4$. Let
$u=u_{0},u_{1},...,u_{k}=v$, $k=\rho(u,v)\geq4$ be a shortest path
connecting the vertices $u$ and $v$. Note that $(u_{1},u_{3})\notin
E(G)$, thus due to (P1), there is $U\in
U_{G}$ with $\{u_{1},u_{3}\}\subseteq U$. Let $z\in U$ be a vertex such that $\{z\}$ is recognizing for $U$. (a) implies that $(u,z)\in E(G)$ and $(u_{4},z)\in E(G)$. Note that $u=u_0,z, u_4,...,u_k=v$ is a path connecting the vertices $u$ and $v$, whose length is smaller than $k=\rho(u,v)$, which is a contradiction. The proof
of the Lemma \ref{SrtuctProperties} is completed. $\square$
\end{proof}

We are ready to present the first main result of the paper, which is a tight lower bound for $c(\mathcal{U}_G)$ in the class of connected graphs $G$ if one is willing to disregard finitely many exceptions.
 
\begin{theorem} \label{MainBound} If $G=(V,E)$ is a connected graph, with $|V(G)|\geq 2$, that
is not isomorphic to $K_{2,2},K_{3,3},K_{4,4}$, then
\begin{equation*}
c(\mathcal{U}_G)\geq \frac{1}{1+|V(G)|-2\sqrt{|V(G)|-1}}.
\end{equation*}
\end{theorem}

\begin{proof}Suppose that there is a $U\in U_{G}$ with $|S_{U}|\geq2$, where $S_{U}$ is a smallest recognizing subset for $U$.
Since $G$ is connected and $|V(G)|\geq 2$, we have $|U|\leq
|V(G)|-1$, thus
\begin{equation*}
c(\mathcal{U}_G)\geq c(U_{0})=\frac{|S_{U_{0}}|}{|U_{0}|}\geq
\frac{2}{|V(G)|-1}\geq \frac{1}{1+|V(G)|-2\sqrt{|V(G)|-1}}.
\end{equation*}

Thus, without loss of generality, we may assume that for each $U\in
U_{G}$ we have $|S_{U}|=1$. Note that if we could prove that in such
graphs
\begin{equation}\label{MainInequality}
|V(G)|\leq1+\biggr(\frac{1+\Delta(G)}{2}\biggr)^{2},
\end{equation}
which is equivalent to
\begin{equation*}
\Delta(G)\geq2\sqrt{|V(G)|-1}-1,
\end{equation*}
then, due to (P3), we would have
\begin{equation*}
c(\mathcal{U}_G)\geq \frac{1}{1+|V(G)|-2\sqrt{|V(G)|-1}},
\end{equation*}
and the proof of the theorem would be completed. Thus, to complete
the proof, it suffices to show that if $G$ is a graph satisfying the
conditions of Theorem \ref{MainBound} and for each $U\in U_{G}$
we have $|S_{U}|=1$, then the inequality (\ref{MainInequality})
holds.

Let $U_{G}=\{U_{1},...,U_{l}\}$, and suppose $Q$ is a maximum clique
of $G$ with $V(Q)=\{v_{1},...,v_{l}\}$, $v_{i}\in
S_{G}(U_{i}),i=1,...,l$ (see (c) of Lemma \ref{SrtuctProperties}). Set:
$V_{0}=V(G)\backslash V(Q)$.

First of all, let us show that each $x\in V_{0}$ has a neighbour in $Q$. Since $G$
is connected and $|V(G)|\geq 2$, there is a $y\in V(G)$ such that
$(x,y)\in E(G)$. Due to (P1), there is a
$U_{y}\in U_{G}$ containing the vertex $y$. Due to (a) and (c) of
Lemma \ref{SrtuctProperties} there is a $z\in V(Q)\cap U_{y}$ such
that $z$ is adjacent to all vertices lying outside $U_{y}$, and
particularly, to $x$.

To complete the proof of the theorem, we need to consider three cases. Note that since $G$ is a connected graph with at least two vertices, we have $l\geq 2$.

Case 1: $l=2$. 

Due to (c) of Lemma
\ref{SrtuctProperties}, $l$ is the size of a maximum clique of $G$, thus $G$ does not contain a triangle.
We claim that $G$ is bipartite. Suppose not, and let $C$ be a
shortest odd cycle of the graph $G$, with
$V(C)=\{z_{1},...,z_{k}\}$,
$E(C)=\{(z_{1},z_{2}),...,(z_{k-1},z_{k}),(z_{k},z_{1})\}$ and
$k\geq5$. Since $C$ is a shortest odd cycle, we have
$(z_{1},z_{4})\notin E(G)$. Due to (P1),
there is $U_{z_{1},z_{4}}\in U_{G}$ containing the vertices $z_{1}$
and $z_{4}$. Let $x\in U_{z_{1},z_{4}}$ be a vertex, such that $\{x\}$ is recognizing for $U_{z_{1},z_{4}}$. (a) of Lemma \ref{SrtuctProperties} implies that $x$ is adjacent to all vertices lying
outside $U_{z_{1},z_{4}}$. Since $z_{2}\notin U_{z_{1},z_{4}}$ and
$z_{3}\notin U_{z_{1},z_{4}}$, we have $(x,z_{2})\in E(G)$ and
$(x,z_{3})\in E(G)$. This is a contradiction since the vertices
$x,z_{2},z_{3}$ induce a triangle.

Thus $G$ is a bipartite graph, and let $(X_{1},X_{2})$ be the
bipartition of $G$, where $V(G)=X_{1}\cup X_{2}$, $X_{1}\cap
X_{2}=\emptyset$. It is clear that $X_{1}\in U_{G}$ and $X_{2}\in
U_{G}$. Since, by assumption $l=2$, we have $U_{G}=\{X_{1},X_{2}\}$.
This and (P1) imply that for each $x_{1}\in
X_{1}$ and $x_{2}\in X_{2}$ we have $(x_{1},x_{2})\in E(G)$. Thus
the graph $G$ is isomorphic to the complete bipartite graph
$K_{m,n}$ for some $m,n$ with $m\geq n$.

Now if $n=1$, then $|V(G)|=m+1$, $\Delta(G)=m$, and therefore
\begin{equation*}
|V(G)|=m+1\leq1+\biggr(\frac{1+m}{2}\biggr)^{2}=1+\biggr(\frac{1+\Delta(G)}{2}\biggr)^{2},
\end{equation*}
thus, we can assume that $n\geq2$. On the other hand, if $m\geq5$,
then $|V(G)|=m+n$, $\Delta(G)=m$, and therefore
\begin{equation*}
|V(G)|=m+n\leq2m\leq1+\biggr(\frac{1+m}{2}\biggr)^{2}=1+\biggr(\frac{1+\Delta(G)}{2}\biggr)^{2},
\end{equation*}
thus, we can assume that $m\leq4$. Since by assumption $G$ is not
isomorphic to $K_{2,2},K_{3,3},K_{4,4}$, then $G$ is either $K_{2,3}$
or $K_{2,4}$ or $K_{3,4}$. It is a matter of direct verification
that these three graphs satisfy the inequality
(\ref{MainInequality}).

Case 2: $l\geq 3$ and $V_{0}$ is an independent set.

(P1) and Lemma \ref{SrtuctProperties} imply
that there is $w\in V(Q)$ such that $(\{w\}\cup V_{0})\in U_{G}$.
Note that all neighbours of $w$ belong to $Q$ and $d(w)=l-1$. Taking
into account that all vertices of $V_{0}$ are adjacent to a vertex
from $Q$, we deduce
\begin{equation*}
|V_{0}|\leq (l-1)(\Delta(G)-(l-1)),
\end{equation*}
and therefore
\begin{equation*}
|V(G)|=|V_{0}|+l\leq
l+(l-1)(\Delta(G)-(l-1))=1+(l-1)((\Delta(G)+1)-(l-1))\leq1+\biggr(\frac{1+\Delta(G)}{2}\biggr)^{2}.
\end{equation*}

Case 3: $l\geq 3$ and $V_{0}$ is not an independent set.

Let $x,y\in V_{0}$ such that $(x,y)\in E(G)$. Assume that $V_{0}=\{x,y,w_1,...,w_k\}$ ($k\geq 0$). Choose a vertex
$z\in V(Q)$. (P1) and (c) of Lemma
\ref{SrtuctProperties} imply that there is $U_{z}\in U_{G}$ such that
$\{z\}$ is recognizing for $U_{z}$. (a) of Lemma
\ref{SrtuctProperties} implies that $(z,x)\in E(G)$ or $(z,y)\in
E(G)$. Taking into account that each vertex of $V_{0}$ has a neighbour in $Q$, we have
\begin{equation*}
|V_{0}|\leq l(\Delta(G)-(l-1))-(l-2).
\end{equation*}Let us show that without loss of generality, we can assume that one has equality above. Suppose that $|V_{0}|< l(\Delta(G)-(l-1))-(l-2)$. Then $|V_{0}|\leq l(\Delta(G)-(l-1))-l+1$, and therefore
\begin{equation*}
|V(G)|=|V_{0}|+l\leq 1+l((\Delta(G)+1)-l)\leq 1+\biggr(\frac{1+\Delta(G)}{2}\biggr)^{2}.
\end{equation*}Thus, we can assume that $|V_{0}|=l(\Delta(G)-(l-1))-(l-2)$. Note that this equality implies:
\begin{itemize}
	\item [(1)] for each $z\in V(Q)$ we have $d(z)=\Delta(G)$;
	\item [(2)] the vertices $w_1,...,w_k$ are of degree one;
	\item [(3)] each vertex $z\in V(Q)$ is adjacent to $\Delta(G)-l$ vertices from $w_1,...,w_k$, and exactly one of $x$ and $y$.
\end{itemize} Since each vertex of $V_{0}$ has a neighbour in $Q$, (2) implies that $V_{0}\backslash \{x\}$ is an independent set, thus due to (P1) and (c) of Lemma
\ref{SrtuctProperties}, there is $z_0\in V(Q)$, such that $((V_{0}\backslash \{x\})\cup \{z_0\})\in U_G$. This particularly means that $z_0$ is not adjacent to any vertex from $\{w_1,...,w_k\}$. This, (1) and (3) imply that:
\begin{equation*}
k=0, |V(G)|=l+2\textrm{ and } \Delta(G)=l.
\end{equation*}
Taking into account that $l\geq 3$, we deduce
\begin{equation*}
|V(G)|=l+2=1+(l+1)\leq
1+\biggr(\frac{1+l}{2}\biggr)^{2}=1+\biggr(\frac{1+\Delta(G)}{2}\biggr)^{2}.
\end{equation*}
The proof of the Theorem \ref{MainBound} is completed.$\square$
\end{proof}

\begin{figure}[h]
\begin{center}
\includegraphics [height=20pc]{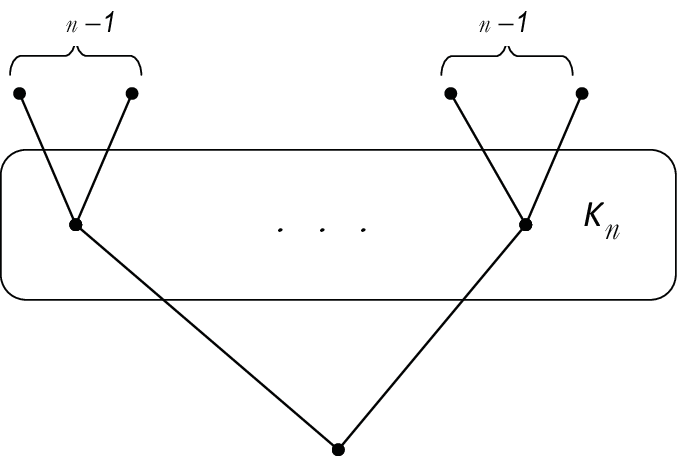}\\
\caption{Example attaining the bound of Theorem
\ref{MainBound}}\label{ExampleBound}
\end{center}
\end{figure}

\begin{remark} There is an infinite sequence of graphs
attaining the bound of the Theorem \ref{MainBound}. For a positive
integer $n$ consider the graph $G$ from Figure \ref{ExampleBound}.
Note that $|V(G)|=1+n^{2},\Delta(G)=2n-1$ and
$c(\mathcal{U}_G)=\frac{1}{n^{2}-2n+2}$.
\end{remark}

\begin{theorem}\label{AllRationals} For any $m,n\in N$ with $1\leq m\leq n$ there is a connected bipartite
graph $G$ such that $c(\mathcal{U}_G)=\frac{m}{n}$.
\end{theorem}
\begin{proof} For any $m,n\in N$ with $1\leq m\leq n$ consider the connected bipartite graph $G$ from the Figure \ref{MNComplexity}.

\begin{figure}[h]
\begin{center}
\includegraphics [scale=0.25]{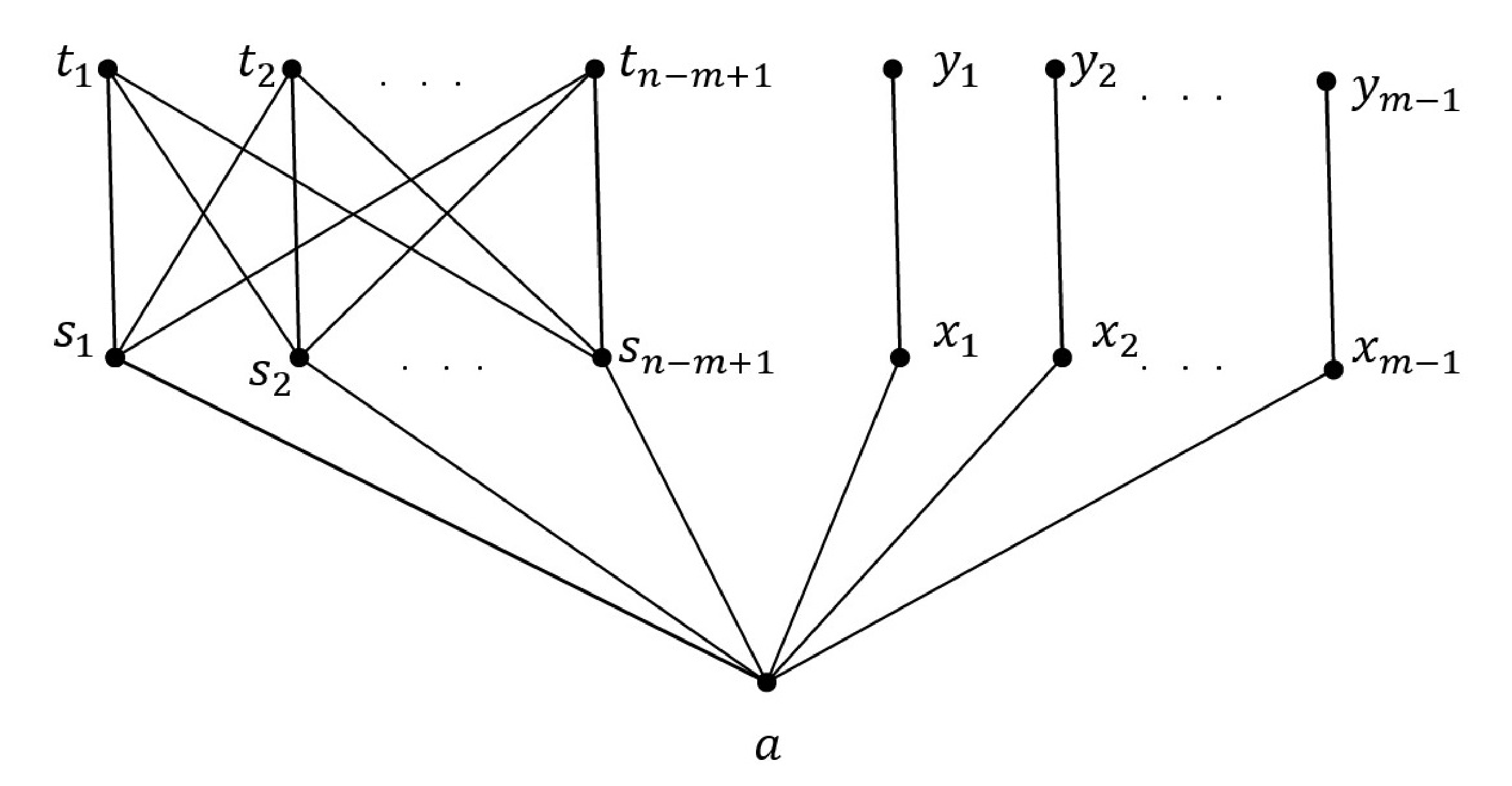}\\
\caption{A graph $G$ with $c(\mathcal{U}_G)=\frac{m}{n}$}\label{MNComplexity}
\end{center}
\end{figure}

Define:
\begin{gather*}
S=\{s_{1},...,s_{n-m+1}\}, T=\{t_{1},...,t_{n-m+1}\},
X=\{x_{1},...,x_{m-1}\}, Y=\{y_{1},...,y_{m-1}\}.
\end{gather*}
Let us show that $c(\mathcal{U}_G)=\frac{m}{n}$. Choose any $U\in U_{G}$. We
will consider two cases.

Case 1: $a\in U$.

Clearly, for each $s\in S, s\notin U$ and for each $x\in X, x\notin
U$, therefore $U=\{a\}\cup T \cup Y$. Lemma
\ref{recognizingCharacter} implies that $S_{U}=\{a\}$ is a smallest recognizing subset for $U$, thus
\begin{equation*}
c(U)=\frac{|S_{U}|}{|U|}=\frac{1}{n+1}.
\end{equation*}

Case 2: $a\notin U$.

It is clear that
\begin{equation}\label{IntersectionXY}
|\{x_{i},y_{i}\}\cap U|=1, \textrm{ for } i=1,...,m-1;
\end{equation}
\begin{equation}\label{IntersectionT}
T\cap U=\emptyset \Leftrightarrow  S\cap U=S \Leftrightarrow
S\subseteq U;
\end{equation}
\begin{equation}\label{IntersectionS}
S\cap U=\emptyset \Leftrightarrow  T\cap U=T \Leftrightarrow
T\subseteq U;
\end{equation}
(\ref{IntersectionXY})-(\ref{IntersectionS}) imply that $|U|=n$.

Now, let $S_U$ be any smallest recognizing subset of $U$. Note that if there is $x_{i}\in U$, then $x_{i}$, with respect
to $y_{i}$ and $U$, satisfies the conditions of the Corollary
\ref{specificVertex}, thus $x_{i}\in S_{U}$. Similarly, if there is
$y_{i}\in U$ then $y_{i}$, with respect to $x_{i}$ and $U$, satisfies
the conditions of the Corollary \ref{specificVertex} (as $a\notin U$),
thus $y_{i}\in S_{U}$.

On the other hand, if $S\subset U$ then due to (\ref{IntersectionT})
$T\cap U=\emptyset$, hence Lemma \ref{recognizingCharacter} implies
that there is $s\in S$ such that $s\in S_U$. Similarly, if $T\subset
U$ then there is $t\in T$ such that $t\in S_U$. This implies that either
there is $s\in S$ such that $(X\cap U)\cup(Y\cap U)\cup
\{s\}\subseteq S_U$ or there is $t\in T$ such that $(X\cap U)\cup(Y\cap U)\cup \{t\}\subseteq
S_U$. Now, it is not hard to see that either $(X\cap U)\cup(Y\cap
U)\cup \{s\}$ or $(X\cap U)\cup(Y\cap U)\cup \{t\}$ is recognizing
for $U$, hence $(X\cap U)\cup(Y\cap U)\cup \{s\}=S_U$ or $(X\cap
U)\cup(Y\cap U)\cup \{t\}=S_U$, and therefore
\begin{gather*}
|S_U|=|X\cap U|+|Y\cap U|+1=m,
\end{gather*}
since, due to (\ref{IntersectionXY}), we have $|X\cap U|+|Y\cap
U|=m-1$. Thus:
\begin{equation*}
c(U)=\frac{|S_{U}|}{|U|}=\frac{m}{n}.
\end{equation*}

The considered two cases imply
\begin{equation*}
c(\mathcal{U}_G)=max\{\frac{1}{n+1},\frac{m}{n}\}=\frac{m}{n}.
\end{equation*}
The proof of the Theorem \ref{AllRationals} is completed.$\square$
\end{proof}

In the end of the section we study the hardness of clutters $\mathcal{U}_T$ arising from trees $T$. Our goal is to try to characterize the class of trees $T$, for which $c(\mathcal{U}_T)=1$. Though we fail to achieve this, we are able to present some non-trivial necessary and sufficient conditions.

\begin{definition}\label{vertextypes}In a tree $T$ a vertex $t\in
V(T)$ is

\begin{enumerate}
\item[(a)] $\alpha$-vertex, if there is $t'\in V(T)$ with $d(t')=1$ and $\rho(t,t')=2$;

\item[(b)] $\beta$-vertex, if it is adjacent to an $\alpha$-vertex, whose all neighbours that differ from $t$, are $\alpha$-vertices;

\item[(c)] $\gamma$-vertex, if it is adjacent to a $\beta$-vertex;

\item[(d)] $\beta$-vertex, if it is adjacent to an $\alpha$-vertex, whose all neighbours that differ from $t$, are $\alpha$ or $\gamma$-vertices;

\item[(e)] $\delta$-vertex, if all its neighbours are $\alpha$ or $\gamma$-vertices;
\end{enumerate}
\end{definition}

\begin{remark} By definition, a vertex of a tree can satisfy
more than one of conditions of Definition \ref{vertextypes}, and
thus be of more than one type.
\end{remark}

\begin{remark}\label{DefExplanation} The definition has a recursive structure, and in (c), in the
definition of a $\gamma$-vertex, a $\beta$-vertex is understood as
one which is defined by (b) or (d). For the sake of clear
explanation and proving the next lemma, we will imagine that our
definition works as a labeling algorithm. The algorithm for its input gets a
tree. During the initialization it labels all $\alpha$-vertices
according to (a) of Definition \ref{vertextypes}. Then at the first
step it labels all $\beta$-vertices and their neighbour
$\gamma$-vertices according to (b) and (c) of Definition
\ref{vertextypes}, respectively. If at the $k^{th}$ step, the
labeling is already done, then in $(k+1)^{th}$ step it labels all
$\beta$-vertices and their neighbour $\gamma$-vertices according to
(d) and (c) of Definition \ref{vertextypes}, respectively. The
process continues until no new vertex receives a label. Finally, in
the last step, the algorithm labels all $\delta$-vertices according
to (e) of Definition \ref{vertextypes} and presents the labeling of
the input tree as the output.
\end{remark}

\begin{remark} By definition, every $\beta$-vertex of a tree is a
$\delta$-vertex, therefore it is natural to introduce the following definition
\end{remark}

\begin{definition}A $\delta$-vertex is called pure, if it is not a
$\beta$-vertex.
\end{definition}

The following lemma explains the essence of Definition
\ref{vertextypes}.

\begin{lemma}\label{EssenceDef}Let $T$ be a tree. Suppose that $U\in U_T$ and
$c(U)=1$. Then:
\begin{enumerate}
\item[(1)] all $\alpha$-vertices do not belong to $U$;

\item[(2)] all $\beta$-vertices belong to $U$;

\item[(3)] all $\gamma$-vertices do not belong to $U$;

\item[(4)] all $\delta$-vertices belong to $U$;

\end{enumerate}
\end{lemma}
\begin{proof} (1) Suppose that $t$ is an $\alpha$-vertex. Then,
due to (a) of Definition \ref{vertextypes}, there is $t'\in V(T)$
with $d(t')=1$ and $\rho(t,t')=2$. If $t\in U$, then the only
neighbour of $t'$, which is also a neighbour of $t$, does not lie in
$U$, hence $t'\in U$ as $U\in U_T$. Now, observe that $U\backslash
\{t'\}$ is a recognizing set for $U$, since it trivially satisfies
the condition of the Lemma \ref{recognizingCharacter}. This implies
that
\begin{gather*}
c(U)\leq \frac{|U|-1}{|U|}<1,
\end{gather*}
which is a contradiction.

(2),(3) We will give a simultaneous proof of (2) and (3) by
induction on $k$, where $k$ is the current step of the labeling
algorithm (Remark \ref{DefExplanation}).

So, assume that $k=1$, $t$ is a $\beta$-vertex and it "became" such
a one due to (b) of Definition \ref{vertextypes}. Let us show that
$t\in U$.

According to (b) of Definition \ref{vertextypes}, there is an
$\alpha$-vertex $t'$, whose all neighbours except $t$, are
$\alpha$-vertices. Due to (1) of Lemma \ref{EssenceDef}, neither
$t'$ nor its $\alpha$-neighbours that differ from $t$, do not belong
to $U$. Since $U\in U_T$, we deduce $t\in U$.

This implies that all $\gamma$-vertices that are
adjacent to a $\beta$-vertex that was labeled in the first step, do
not belong to $U$. Thus (2) and (3) are true for $k=1$.

Now, assume that (2) and (3) are true for vertices which receive
their labels in the steps up to $k$. Consider a $\beta$-vertex $t$
which gets its label according to (d) of Definition
\ref{vertextypes} in the $(k+1)^{th}$ step of the labeling
algorithm. Let us show that $t\in U$.

According to (d) of Definition \ref{vertextypes}, there is an
$\alpha$-vertex $t'$, whose all neighbours except $t$, are $\alpha$
or $\gamma$-vertices, which have received their labels earlier than
the $(k+1)^{th}$ step. Due to the induction hypothesis and (1) of
Lemma \ref{EssenceDef}, neither $t'$ nor its $\alpha$ or
$\gamma$-neighbours that differ from $t$, belong to $U$. Since $U\in
U_T$, we deduce $t\in U$.

This implies that all $\gamma$-vertices that are adjacent to a
$\beta$-vertex that was labeled in the $(k+1)^{th}$ step, do not
belong to $U$. Thus (2) and (3) are true for $k+1$ and the proof is completed.

(4) If $t$ is a $\delta$-vertex, then due to (e) of Definition
\ref{vertextypes}, and (1) and (3) of Lemma \ref{EssenceDef}, all
the neighbours of $t$ do not belong to $U$, hence $t\in U$ as $U\in
U_T$.

The proof of the Lemma \ref{EssenceDef} is completed.$\square$
\end{proof}

The proved lemma implies the following necessary
condition for a tree $T$ to satisfy $c(\mathcal{U}_T)=1$.

\begin{corollary}\label{TreesNecessaryCondition}If $T$ is a tree
with $c(\mathcal{U}_T)=1$, then:
\begin{enumerate}
\item[(a)] there is no $\alpha$ or $\gamma$-vertex, which is also a $\beta$ or a $\delta$-vertex;

\item[(b)] each $\delta$-vertex $t$ is adjacent to an $\alpha$ or a $\gamma$-vertex, that has a neighbour that is different from $t$ and which is neither a $\beta$ nor a $\delta$-vertex.
\end{enumerate}
\end{corollary}
\begin{proof} (a) is clear.

(b) On the opposite assumption, consider a $\delta$-vertex $t$, whose
all neighbours are $\alpha$ or $\gamma$-vertices ((e) of Definition
\ref{vertextypes}), and whose every neighbour that is different from $t$ is adjacent to a
$\beta$ or a $\delta$-vertex. Due to
Lemma \ref{EssenceDef}, the vertex $t$ and these $\beta$ or
$\delta$-vertices lying on a distance two from $t$ belong to any
$U\in U_T$ with $c(U)=1$. Now, note that $U\backslash \{t\}$ is a
recognizing set for $U$, since it trivially satisfies the condition
of the Lemma \ref{recognizingCharacter}. This implies that
\begin{gather*}
c(U)\leq \frac{|U|-1}{|U|}<1,
\end{gather*}
which is a contradiction.$\square$
\end{proof}

\begin{theorem}\label{TreesSuff}If a tree $T$ contains neither a $\beta$ nor
a pure $\delta$-vertex, then for each $u\in V(T)$ with $d(u)=1$
there is $U\in U_T$ with $c(U)=1$ and $u\in U$.
\end{theorem}
\begin{proof}Unfortunately, the proof of existence of such $U\in U_T$ is not
easy. This is the main reason that we will give an algorithmic
construction of such $U\in U_T$.

Given $u\in V(T)$ with $d(u)=1$, we will assume that $T$ is
represented as a tree rooted at $u$.\\

Step 0:
\begin{equation*}
U:=\{u\},Spec:=\{\textrm{the neighbours of } u\}
\end{equation*}
Consider the sets $B_1,...,B_k$ of vertices lying at a distance
three from $u$, where it is assumed that the vertices of $B_j,1\leq
j\leq k$ are adjacent to the same vertex. Let $List$ be a list
comprised of the sets $B_1,...,B_k$. Note that since $T$ does not
contain a $\beta$-vertex, we have that all of $B_1,...,B_k$ contain
a
non-$\alpha$ vertex.\\

Step 1: while $List\neq \emptyset$

remove the first element $B$ of $List$.

Define $A=\{v\in B:v \textrm{ is not a }\alpha \textrm{-vertex} \}$

$A'=\{v\in A:\textrm{ all children of }v \textrm{ are } \alpha
\textrm{-vertices} \}$\\

Case 1: $A'\neq \emptyset$

$U:=U\cup A'$

Add all children of vertices from $A'$ (which are $\alpha$
-vertices, by definition) to the set $Spec$.

Note that, by definition of $A'$, for each $w\in A\backslash A'$ the
set $B_w$, which is the set of children of $w$, contains a
non-$\alpha$ vertex. Moreover, for each $z\in A'$ if we consider the
sets $B_{z_1},...,B_{z_s}$ of vertices lying at a distance three
from $z$ (the vertices of $B_{z_j},1\leq j\leq s$ are adjacent to
the same vertex), then since $T$ contains no $\delta$-vertex, each
of these sets contains a non-$\alpha$ vertex.

Add all $B_w,B_{z_1},...,B_{z_s}$ to $List$;\\

Case 2: $A'=\emptyset$

Take any $w\in A.$

$U:=U\cup \{w\};$ add the parent $x$ of $w$ to the set $Spec$.

Note that $A'=\emptyset$ implies that for each $y\in B\backslash
\{w\}$ the set $B_w$ of children of $y$ contains a non-$\alpha$
vertex. On the other hand, since $T$ contains no $\beta$-vertex,
then for each $z\in B\backslash A$ the set $B_z$ of children of $z$
contains a non-$\alpha$ vertex.

Add all $B_w,B_{z}$ to $List$;

Consider the sets $B_i$ of vertices lying at a distance three from
$w$, where it is assumed that $B_i$ is the set of children of $z_i$.\\

Case 2.1: $B_i$ contains a non-$\alpha$ vertex;

Add $B_i$ to $List$;\\

Case 2.2: All vertices of $B_i$ are $\alpha$-vertices;

$U:=U\cup \{z_i\};$ $Spec:=Spec\cup B_i;$

Consider the sets $B_{z_1}^{(i)},...,B_{z_s}^{(i)}$ of vertices
lying at a distance three from $z$, where we assume that
$B_{z_j}^{(i)}$ coincides with the set of children of a vertex
${z_j}^{(i)}$. Since $T$ contains no $\delta$-vertex, then each
$B_{z_j}^{(i)}$ contains a non-$\alpha$ vertex.

Add $B_{z_1}^{(i)},...,B_{z_s}^{(i)}$ to $List$;

The description of the algorithm is completed.\\

Let us note that if the algorithm cannot choose the set $A$ then the
last vertex from which it is impossible to choose a vertex lying on
a distance three, is either a pendant vertex, which has a specific
vertex in the set $Spec$, or is a vertex that is adjacent to a
pendant vertex, and this pendant vertex will be the specific vertex
for it.

It can be easily seen that the algorithm constructs a maximal
independent set $U$ of $T$ containing the vertex $u$. The
construction of the set $Spec$ implies that each vertex $v\in U$ has
a specific neighbour in $Spec$, that is, a neighbour, which is not
adjacent to any other vertex of $U$. This and Corollary
\ref{specificVertex} imply that the hardness of $U$ is one. The
proof of Theorem \ref{TreesSuff} is completed.$\square$
\end{proof}

\begin{figure}[h]
\begin{center}
\includegraphics [height=30pc]{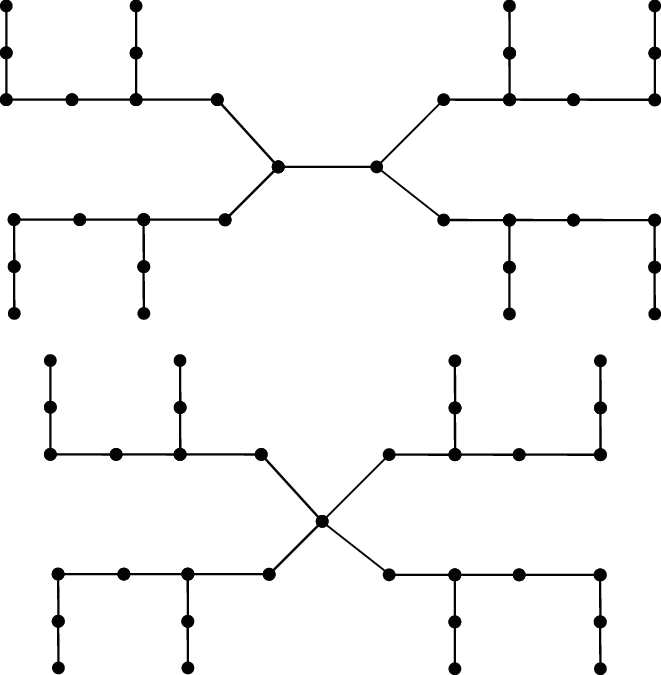}\\
\caption{Trees with pure $\delta$-vertices, without $\beta$-vertices
}\label{ExampleTreesGamma}
\end{center}
\end{figure}

\begin{figure}[h]
\begin{center}
\includegraphics [height=30pc]{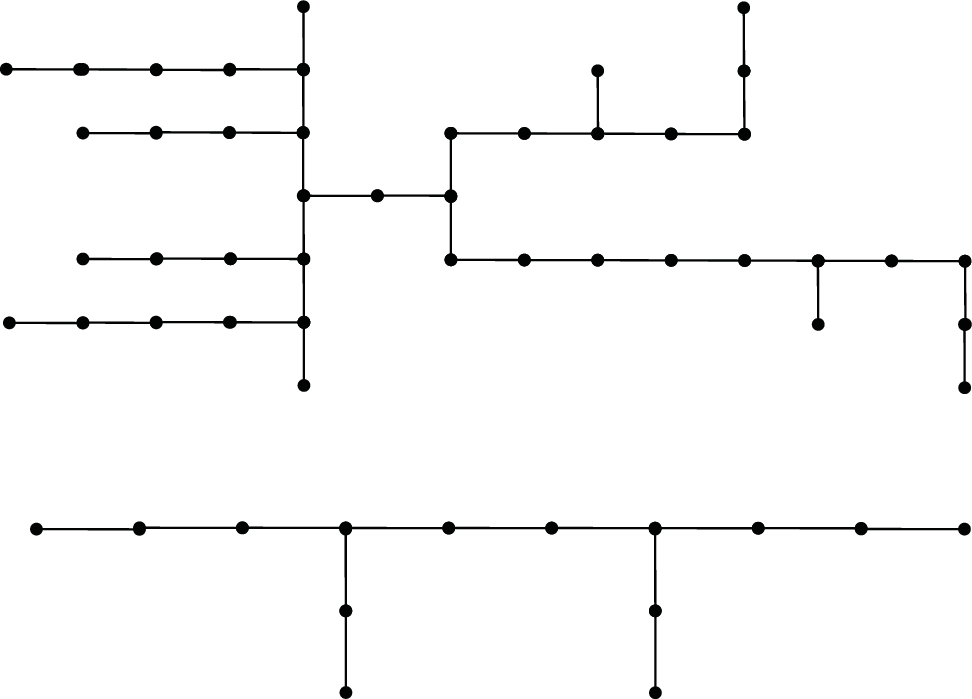}\\
\caption{Trees with $\beta$-vertices, without pure $\delta$-vertices
}\label{ExampleTreesBetta}
\end{center}
\end{figure}

\begin{remark} The Theorem \ref{TreesSuff} presents merely a
sufficient condition. The trees from Figure \ref{ExampleTreesGamma}
contain a pure $\delta$-vertex, do not contain a $\beta$ vertex, and
nevertheless, the first of them has a hardness that is less than
one, while the second one is of hardness one. On the other hand,
the trees from Figure \ref{ExampleTreesBetta} contain a
$\beta$-vertex, do not contain a pure $\delta$ vertex, and
nevertheless, the first of them has a hardness that is less than
one, while the second one is of hardness one.
\end{remark}

\section{The hardness of $\mathcal{M}_G$}

Below we investigate the hardness of the clutter $\mathcal{U}_G$ in the class of line graphs $G$. This class is interesting not only for its own sake, but also for
its connection with another clutter related to graphs. Taking into
account, that the clutter $U_G$ of a line graph $G$ coincides with
the clutter $\mathcal{M}_H$ of some graph $H$, in this sections we
will directly work with the latter clutter without remembering that
it was originated from a line graph.

\subsection{Structural Lemmas}

\begin{lemma}
\label{basiclemma} Assume that $H\in \mathcal{M}_G$ and let $S_H$ be any smallest recognizing subset of $H$. Then:

\begin{enumerate}
\item The vertices of $V(H\backslash S_H)$ can only be connected to the
vertices of $V(S_H)$.

\item Each edge in $S_H$ has at least one endpoint connected to a vertex not
in $V(S_H)$.
\end{enumerate}
\end{lemma}

\begin{proof}
Let $e=(u,v)$ be an edge in $H\backslash S_H$. Let us first prove
that both $u$ and $v$ are not connected to vertices, which are not
covered by $H$. If this is not true, then without loss of generality
we may assume that $\exists p\in V(G)\backslash V(H)$, such that
$(p,u)\in E(G)$. $H\cup \{(p,u)\}\backslash \{(u,v)\}$ is a maximal
matching containing $S_H$. This contradicts the definition of $S_H$.

We have proven that the vertices of $V(H\backslash S_H)$ can only be
connected to the vertices of $V(H)$.

Now, if there are there are vertices $\{u_1,u_2,u_3,u_4\}$, such
that
\begin{equation*}
(u_1,u_2), (u_3,u_4) \in H\backslash S_H
\end{equation*}
and $(u_2,u_3)\in E(G)$, then there is a maximal matching that
contains $(H\backslash \{(u_1,u_2),(u_3,u_4)\})\cup\{(u_2,u_3)\}$.
That maximal matching is different from $H$ and contains $S_H$. This
is a contradiction proving point 1.

If the statement of point 2 does not take place for an edge $e$,
then every maximal matching, which contains $S_H\backslash \{e\}$
also contains $S_H$. Thus $H$ is the only maximal matching, which
contains $S_H\backslash \{e\}$, and consequently $S_H$ is not a
minimum subset of $H$ with this property. The contradiction proves
point 2.$\square$
\end{proof}

\begin{lemma}\label{minimumlemma}
Suppose $H$ is a smallest maximal matching in $G$ and $e\in H$. The
endpoints of $e$ cannot be connected to endpoints of different edges
of $H\backslash S_H$, where $S_H$ is any smallest recognizing subset of $H$.
\end{lemma}

\begin{proof}
Let $(u,v)$ be an edge in $S_H$. If there are edges $(u_1, v_1)$ and
$(u_2,v_2)$ from $H\backslash S_H$, such that $u$ is connected to
$u_1$ and $v$ is connected to $u_2$, then $H$ is not a smallest
maximal matching since the cardinality of
\begin{equation*}H\cup\{(u,u_1),(v,u_2)\}\backslash\{(u,v),(u_1,v_1),(u_2,v_2)\}\end{equation*}
 is less than that of $H$.$\square$
\end{proof}

Also, recall the following result \cite{Lov,Sumner}:

\begin{lemma}
\label{lovEx}If $G$ is a connected graph, whose every maximal
matching is a perfect matching, then $G$ is either $K_{2n}$ or
$K_{n,n}$.
\end{lemma}

\subsection{A lower bound for hardness}

\begin{figure}[h]
\begin{center}
\includegraphics [height=21pc]{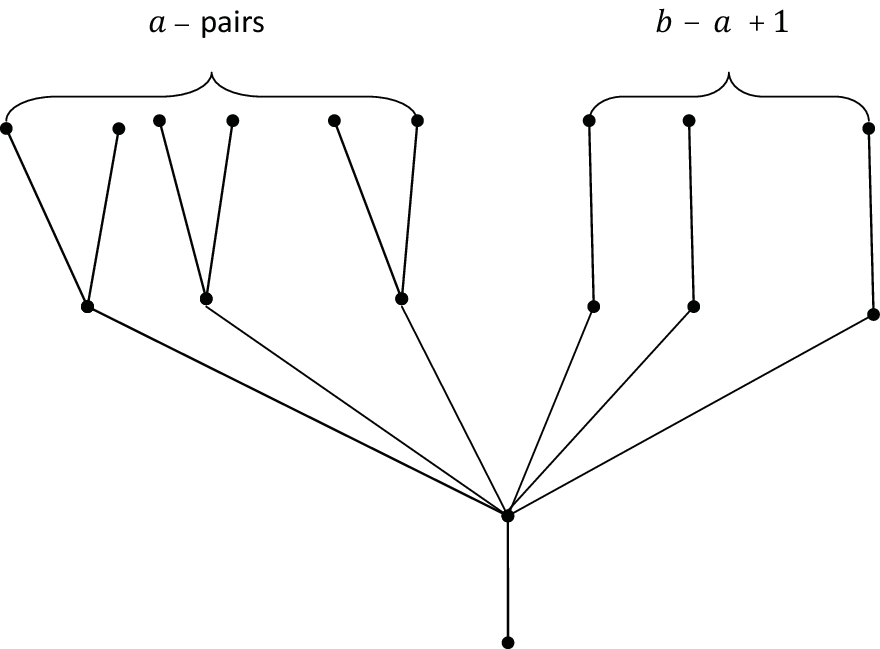}\\
\caption{A graph $G$ with $c(\mathcal{M}_G)=r$}\label{MNComplexityTrees}
\end{center}
\end{figure}

Note that the hardness of $\mathcal{M}_G$ for disconnected graphs $G$ does not have a
lower bound better than zero. For instance, for the graph $K$ that consists
of a single matching, we have $c(\mathcal{M}_K)=0$. Moreover, it can be
shown that for every rational number $r$, $0\leq r\leq 1$ there
exists a graph with hardness $r$. To construct one just consider
the graph $G_{r}$ from Figure \ref{MNComplexityTrees}, where we
assume that $r=\frac{a+1}{b+1}$.

The following theorem proves a tight lower bound for the
hardness of $\mathcal{M}_G$ in the class of connected graphs $G$. Before we move on, let us note that the bound given in the theorem below, is significantly better
than the one that Theorem \ref{MainBound} provides.

\begin{theorem}\label{ComplexityBoundMG}
For every connected graph $G$ with $|V(G)|>4$, we have $c(\mathcal{M}_G)\geq
\frac2{|V(G)|-2}$.
\end{theorem}

\begin{proof}
Let $H$ be a smallest maximal matching of $G$, and let $S_H$ be any smallest recognizing subset of $H$. If
$|H|<\lfloor|V|/2\rfloor$ then $|H|\leq\frac{|V|-2}2$ and
\begin{equation}
c(\mathcal{M}_G)\geq
c(H)=\frac{|S_H|}{|H|}\geq\frac1{(|V|-2)/2}=\frac2{|V|-2}
\end{equation}

If $|H|=\lfloor|V|/2\rfloor$, then there are two cases.
\begin{itemize}
\item $|V|$ is even. Since $H$ is a smallest maximal matching,
every maximal matching of $G$ is a perfect matching. Due to Lemma
\ref{lovEx}, $G$ is isomorphic to either $K_{2n}$ or
$K_{n,n}$($n=|V|/2>2$). For these graphs
\begin{equation*}c(\mathcal{M}_G)=\frac{n-1}n>\frac1{n-1}=\frac2{|V|-2}\end{equation*}
\item $|V|$ is odd. If $|S_H|\geq2$ then
\begin{equation}\label{S3equ}
c(\mathcal{M}_G)\geq2/|H|=4/(|V|-1)\geq2/(|V|-2) \end{equation}
 Assume $S_H=\{(u,v)\}$. Lemma \ref{basiclemma} implies that either $|H|=2$($|V|=5$) or all the vertices of
$V(H\backslash S_H)$ are connected to only one of the endpoints of
$(u,v)$. Without loss of generality we may assume that they are
connected to $u$.

If $|V|=5$, there are only a few graphs for which it is possible to
have $|H|=2$ and $|S_H|=1$. All these graphs $G$ can be easily checked
to satisfy $c(\mathcal{M}_G)=1$.\\Assume $|H|\geq3$. Let $w$ be the
vertex, which is not covered by $H$. If $w$ is connected to $v$ then
due to 1 of Lemma \ref{basiclemma}, we have
$|S_{H\cup\{(v,w)\}\backslash\{(u,v)\}}|> 1$, since all the edges of
$H\cup\{(v,w)\}\backslash\{(u,v)\}$ are connected to $u$. As a
result, according to (\ref{S3equ}),
\begin{equation*}
c(\mathcal{M}_G)>2/(|V|-2).
\end{equation*}

If $w$ is connected to $u$, take an edge $(u_1,v_1)\in H$ such that
$(u,u_1)\in E$. $(H\cup\{(u,u_1)\})\backslash\{(u,v),(u_1,v_1)\}$ is
a maximal matching with a smaller cardinality than $H$. Thus $H$ is
not smallest and this case is impossible.
\end{itemize}
The proof is now completed.$\square$
\end{proof}

Figure \ref{MNComplexityTrees} with $a=0$ illustrates that the bound
achieved in the previous theorem is tight. The depicted graph $G$
contains $2(b+2)$ vertices and it satisfies $c(\mathcal{M}_G)=1/(b+1)$,
therefore
\begin{equation*}
c(\mathcal{M}_G)=\frac2{|V(G)|-2}.
\end{equation*}%

\subsection{Bounds for $c(\mathcal{M}_G)$ in the class of regular graphs $G$}

For regular graphs $G$, it is possible to find lower bounds for $c(\mathcal{M}_G)$ that do not depend on the number of edges in those
graphs.

\begin{theorem}
\label{1/2theorem} For an $r$-regular graph $G$ with $r>1$ $c(\mathcal{M}_G)\geq \frac{1}{2%
}$.
\end{theorem}

\begin{proof}
Take any $H\in \mathcal{M}_G$, and let $S_H$ be any smallest recognizing subset of $H$. Let $E_1$ be the set of edges that
connect $V(S_H)$ with $V(H\backslash S_H)$, $E_2$ be the set of
edges that connect $V(S_H)$ with $V(G)\backslash V(H)$, and $E_3$ be
the set of edges in the spanning subgraph of $V(S_H)$, not including
the edges from $S_H$.\\According to point 1 of Lemma
\ref{basiclemma}, all the vertices of $V(H\backslash S_H)$ are only
connected to the vertices of $V(S_H)$. Therefore,
\begin{gather*}
2|S_H|(r-1)=|V(S_H)|(r-1)=\sum_{v\in
V(S_H)}(d(v)-1)=|E_1|+|E_2|+2|E_3|\geq |E_1|\\
=\sum_{v\in V(H\backslash S_H)}(d(v)-1)=(r-1)|V(H\backslash
S_H)|=2|H\backslash S_H|(r-1).
\end{gather*}
Since $r\neq 1$, we have $|S_H|\geq
|H\backslash S_H|$, thus $c(H)=|S_H|/|H|\geq\frac12$, and therefore
$c(\mathcal{M}_G)\geq\frac12$.$\square$
\end{proof}

\begin{corollary}
If $G$ is a regular graph and $c(\mathcal{M}_G)=\frac12$ then for every maximal
matching $H$, $c(H)=\frac12$.
\end{corollary}

\begin{corollary}
\label{perfCor} If $G$ is a regular graph and $c(\mathcal{M}_G)=\frac12$ then
every maximal matching is a perfect matching.
\end{corollary}

\begin{proof}
Since $c(H)=\frac12$, we have that $|S_H|=|H\backslash
S_H|$, and therefore $E_2=\emptyset$. Now suppose there is vertex $v$,
which is not covered by $H$. As $H$ is maximal, it covers all the
neighbors of $v$. Due to 1 of Lemma \ref{basiclemma}, these
neighbors cannot belong to $V(H\backslash S_H)$; consequently, they
belong to $V(S_H)$. This contradicts with $E_2$ being empty.$\square$
\end{proof}

\begin{corollary}
The hardness of $\mathcal{M}_G$ for a connected regular graph $G$ equals $\frac12$ if and only
if $G$ is $K_4$ or $K_{2,2}$.
\end{corollary}

\begin{proof}
It is not hard to see that $c(\mathcal{M}_{K_{2n}})=c(\mathcal{M}_{K_{n,n}})=\frac{n-1}n$. This
said, the corollary follows from Lemma \ref{lovEx} and Corollary
\ref{perfCor}.$\square$
\end{proof}

The following theorem shows that there exist better bounds for the
complexities of $\mathcal{M}_G$ for regular graphs $G$, if we do not consider graphs of small
regularity.

\begin{theorem}For an $r$-regular graph $G$, we have
\begin{enumerate}
\item[(a)]If $r>4$ then $c(\mathcal{M}_G)\geq\frac23$;

\item[(b)]If $r=4$ then $c(\mathcal{M}_G)>\frac35$.
\end{enumerate}
\end{theorem}

\begin{proof}(a)
Due to Lemma \ref{basiclemma}, for each $(u,v)\in S_H$ there are two options:
\begin{itemize}
\item $u$ and $v$ can be connected to the endpoints of only one edge
from $H\backslash S_H$.
\item $u$ is not connected to any vertex covered by $H\backslash S_H$ and $v$
may be connected to any number of endpoints of edges from
$H\backslash S_H$.
\end{itemize}
Therefore, the edges of $S_H$ are divided into two categories. Let
$A$ denote the set of edges of the first category, and $B$ the set
of the edges of the second category. If an edge from $S_H$ falls in
both categories, we will consider it to be in category $A$ and not
$B$.

Retaining the notations of the proof of Theorem \ref{1/2theorem}, we
have $|E_1|=2(r-1)|H\backslash S_H|$. The endpoints of each edge in
category $A$ are the endpoints of at most 4 edges from $|E_1|$. The
endpoints of each edge in category $B$ are the endpoints of at most
$r-1$ edges of $E_1$. This implies:
\begin{equation*}
|E_1|\leq 4|A| +(r-1)|B|=(r-1)|S_H|-(r-5)|A|\leq(r-1)|S_H|.
\end{equation*}

We got that $2|H\backslash S_H| \leq|S_H|$, hence, $c(\mathcal{M}_G)\geq
c(H)=\frac{|S_H|}{|H|}\geq\frac23$.

(b)We will assume that $G$ is connected, because the case of
disconnected graphs easily follows from the case of connected
graphs. Choose any smallest maximal matching $H$ of $G$.

Note that (2) of Lemma \ref{basiclemma} implies that if $e=(u,v)\in
A$ then $u_1=v_1$ or $(u_1,v_1)\in H\backslash S_H$. Moreover,
$S_H=A\cup B,A\cap B=\emptyset$, and

\begin{equation*}
|E_1|=2(r-1)|H\backslash S_H|=6|H\backslash S_H|.
\end{equation*}
The endpoints of each edge in category $A$ are the endpoints of at
most $4$ edges from $E_1$, while the endpoints of each edge in
category $B$ are the endpoints of at most $3$ edges of $E_1$. This
implies:
\begin{equation*}
6|H\backslash S_H|=|E_1|\leq 4|A| +3|B|\leq 4|A| + 4|B| =4|S_H|,
\end{equation*}
or
\begin{equation*}
6|H|\leq 10|S_H|,
\end{equation*}
and therefore
\begin{equation}\label{Eq35}
c(H)=\frac{|S_H|}{|H|}\geq\frac35.
\end{equation}

Now, we claim that $c(H)>\frac35$. If $c(H)=\frac35$ then
\begin{equation*}
|E_1|=4|S_H|=4|A|,
\end{equation*}

and therefore $B=\emptyset$. This implies that for each $e=(u,v)\in
S_H$ there is exactly one $f=(u_1,v_1)\in H\backslash S_H$ such that
\begin{equation*}
\{(u,u_1),(u,v_1),(v,u_1),(v,v_1),\}\subseteq E_1.
\end{equation*}
The uniqueness of $f$ follows from Lemma \ref{minimumlemma}. Note
that this correspondence is one-to-one since $G$ is $4$-regular and an edge from
$H\backslash S_H$ cannot be connected to two different edges from $A$. Thus,
\begin{equation*}
|H|=|S_H|,
\end{equation*}
and
\begin{equation*}
c(H)=\frac{|S_H|}{|H|}=\frac12<\frac35,
\end{equation*}
contradicting (\ref{Eq35}). The proof is now completed.$\square$
\end{proof}

Note that the bound from (a) of the previous theorem is reachable, since $K_6$ is a $5$-regular graph
with $c(\mathcal{M}_{K_6})=\frac23$.

Our interest toward the hardness and particularly, the hardness
of clutters arising from regular graphs was motivated by the following

\begin{conjecture}If $G$ is a connected regular graph with
$c(\mathcal{M}_G)<1$, then $G$ is either isomorphic to $C_7$, or there is
$n,n\geq1$ such that $G$ is isomorphic either to  $K_{n,n}$ or to
$K_{2n}$, where $C_7$ is the cycle of length seven.
\end{conjecture}

In some sense, our conjecture states that all regular structures are
"hard" except some "uninteresting" cases.

\section{Computational complexity results for hardness}

The aim of this section is the investigation of some problems that
are related to the algorithmic computation of the hardness of $\mathcal{U}_G$.

We start with a problem that is related to finding a recognizing set
for a given maximal independent set.\\

\textbf{Problem 1:}

\textbf{Condition: }Given a graph $G$, $U\in U_G$ and a positive
integer $k$.

\textbf{Question: }Is there a recognizing set $U'\subseteq U$ for
$U$ with $|U'|=k$?

\begin{theorem} The \textbf{Problem 1} is $NP$-complete already for bipartite graphs.
\end{theorem}
\begin{proof}Lemma \ref{recognizingCharacter} implies that the \textbf{Problem
1} belongs to the class $NP$. To show the completeness of the
problem, we will reduce the classical \textbf{Set Cover} problem to
our problem restricted to bipartite graphs. Recall that the
\textbf{Set Cover} is formulated as follows (\cite{GareyJohnson}):\\

\textbf{Problem: Set Cover}

\textbf{Condition: }Given a set $A=\{a_1,...,a_n\}$, a family
$\mathcal{A}=\{A_1,...,A_m\}$ of subsets of the set $A$ with
$A_1\cup...\cup A_m=A$, and a positive integer $l,l\leq m$.

\textbf{Question: }Are there $A_{i_1},...,A_{i_l}\in \mathcal{A}$
with $A_{i_1}\cup...\cup A_{i_l}=A$?\\

For an instance $I$ of \textbf{Set Cover} consider the graph
$G_I=(V,E)$, where
\begin{gather*}
V=\{a_1,...,a_n,A_1,...,A_m\}, E=\{(a_i,A_j):a_i\in A_j,1\leq i\leq
n,1\leq j\leq m\}.
\end{gather*}
Note that $G_I$ is bipartite. Consider the set $U=\{A_1,...,A_m\}$.
Since $A_1\cup...\cup A_m=A$, we have $U\in U_{G_I}$.

It can be easily verified that the set $U$ has a recognizing subset
comprised of $l$ elements if and only if there are
$A_{i_1},...,A_{i_l}\in \mathcal{A}$ with $A_{i_1}\cup...\cup
A_{i_l}=A$. The proof of the theorem is completed.$\square$
\end{proof}

Now, we are turning to the investigation of the computation of
$c(\mathcal{U}_G)$. Consider the following\\

\textbf{Problem 2:}

\textbf{Condition: }Given a graph $G$ and positive integers $k,m$
with $1\leq k\leq m$.

\textbf{Question: }Does the inequality $c(\mathcal{U}_G)\leq \frac{k}{m}$ hold?

\begin{theorem} The \textbf{Problem 2} is $NP$-hard already for bipartite graphs.
\end{theorem}
\begin{proof}We will reduce \textbf{Set Cover} to
our problem restricted to bipartite graphs. Given an instance $I$ of \textbf{Set Cover}, consider the
graph $G_I=(V,E)$, where
\begin{gather*}
V=\{A_1,...,A_m\}\cup \{a_{i}^{(k)}:1\leq i\leq n,1\leq k\leq
(n+m)^{2}\},\\
E=\{(a_{i}^{(k)},A_j):a_i\in A_j,1\leq i\leq n,1\leq j\leq m,1\leq
k\leq (n+m)^{2}\}.
\end{gather*}
Note that $G_I$ is bipartite. Let us show that
\begin{equation*}
c(\mathcal{U}_{G_I})=\frac{l_{min}}{m},
\end{equation*}
where $l_{min}$ denotes the size of minimum cover of $A$, that is, the minimum number $l_{min}$ for which there are
$A_{i_1},...,A_{i_{l_{min}}}\in \mathcal{A}$
with $A_{i_1}\cup...\cup A_{i_{l_{min}}}=A$.

Choose any $U\in U_{G_I}$. We will consider two cases.

Case 1: $U=\{A_1,...,A_m\}$.

Lemma \ref{recognizingCharacter} and the definition of $G_I$ imply
that $|S_U|=l_{min}$, therefore
\begin{equation*}
c(U)=\frac{l_{min}}{m}.
\end{equation*}

Case 2: $U\neq \{A_1,...,A_m\}$.

Suppose that $U\cap \{A_1,...,A_m\}=\{A_{i_1},...,A_{i_r}\}$. Since
$U\neq \{A_1,...,A_m\}$, we imply that $A_{i_1}\cup...\cup
A_{i_r}\neq A$. Assume that there are $r',r'\geq1$ elements of $A$
that do not belong to either of $A_{i_j}$'s. Note that all
$r'(n+m)^{2}$ copies of these $r'$ elements belong to $U$, and
\begin{equation*}
|U|=r+r'(n+m)^{2}.
\end{equation*}
On the other hand, if we consider the set $U'\subseteq U$, where
\begin{equation*}
U'=\{A_{i_1},...,A_{i_r}\}\cup \{a_{i}^{(1)}:a_{i}\textrm{ does not
belong to either of } A_{i_j}\textrm{'s}\},
\end{equation*}
then, according to Lemma \ref{recognizingCharacter}, this would be a
recognizing set for $U$, therefore
\begin{gather*}
c(U)=\frac{|S_{U}|}{|U|}\leq\frac{|U'|}{|U|}=\frac{r+r'}{r'(n+m)^{2}}\leq\frac{n+m}{(n+m)^{2}}=
\frac{1}{n+m}<\frac{1}{m}\leq \frac{l_{min}}{m}.
\end{gather*}

The considered two cases imply $c(\mathcal{U}_{G_I})=\frac{l_{min}}{m}$. Now, it is not hard to verify that in the instance $I$ of
\textbf{Set Cover}, there is a cover of length $l$, if and only if $l_{min}\leq l$, which is equivalent to $c(\mathcal{U}_{G_I})\leq \frac{l}{m}$. The proof of the theorem is completed.$\square$
\end{proof}

In the end of the paper, let us note that we have failed to achieve similar results for the clutters $\mathcal{M}_G$. We leave the investigation of the computational complexity of the calculation of $c(\mathcal{M}_G)$ as a research problem.

\begin{acknowledgement}
We thank the referees for their comments that helped us to improve the presentation of the paper.
\end{acknowledgement}

\end{document}